\documentclass[12pt]{iopart}

\usepackage{rotating} 
\usepackage{graphicx}
\usepackage{subfig}

\begin{document}

\title[]{Antiferromagnetic phase transition in $Cr_{2}As$ via anisotropy of exchange interactions}

\author{V.I. Valkov$^1$, A.V. Golovchan$^{1,2}$,Upali Aparajita$^{3}$, Oleksiy Roslyak$^4$}

\address{$^1$Donetsk Institute for Physics and Engineering, Donetsk, Ukraine\\
       $^2$Donetsk National University, Donetsk, Ukraine\\
       $^3$BMCC, CUNY, New York, USA\\
       $^4$Fordham University, New York, USA\\}
\ead{uaparajita@bmcc.cuny.edu}
\vspace{10pt}
\begin{indented}
\item[]September 2018
\end{indented}
\begin{abstract}
The electronic structure of anti-ferromagnetic $Cr_{2}As$ is investigated. Anisotropy of exchange interactions between chrome sub-lattices is determined ($J^{X}(Cr_{I}-Cr_{II}) =4.77 meV,J^{Y}(Cr_{I}-Cr_{II}) =-6.36 meV$). The behavior of exchange integrals from magnetic structure is analyzed. 
\end{abstract}
\vspace{2pc}

\submitto{Advances in Materials Science and Engineering}
\maketitle
\normalsize

\section{Introduction}

\par
\begin{figure}[h!]
\centering
\includegraphics[width=0.6\textwidth]{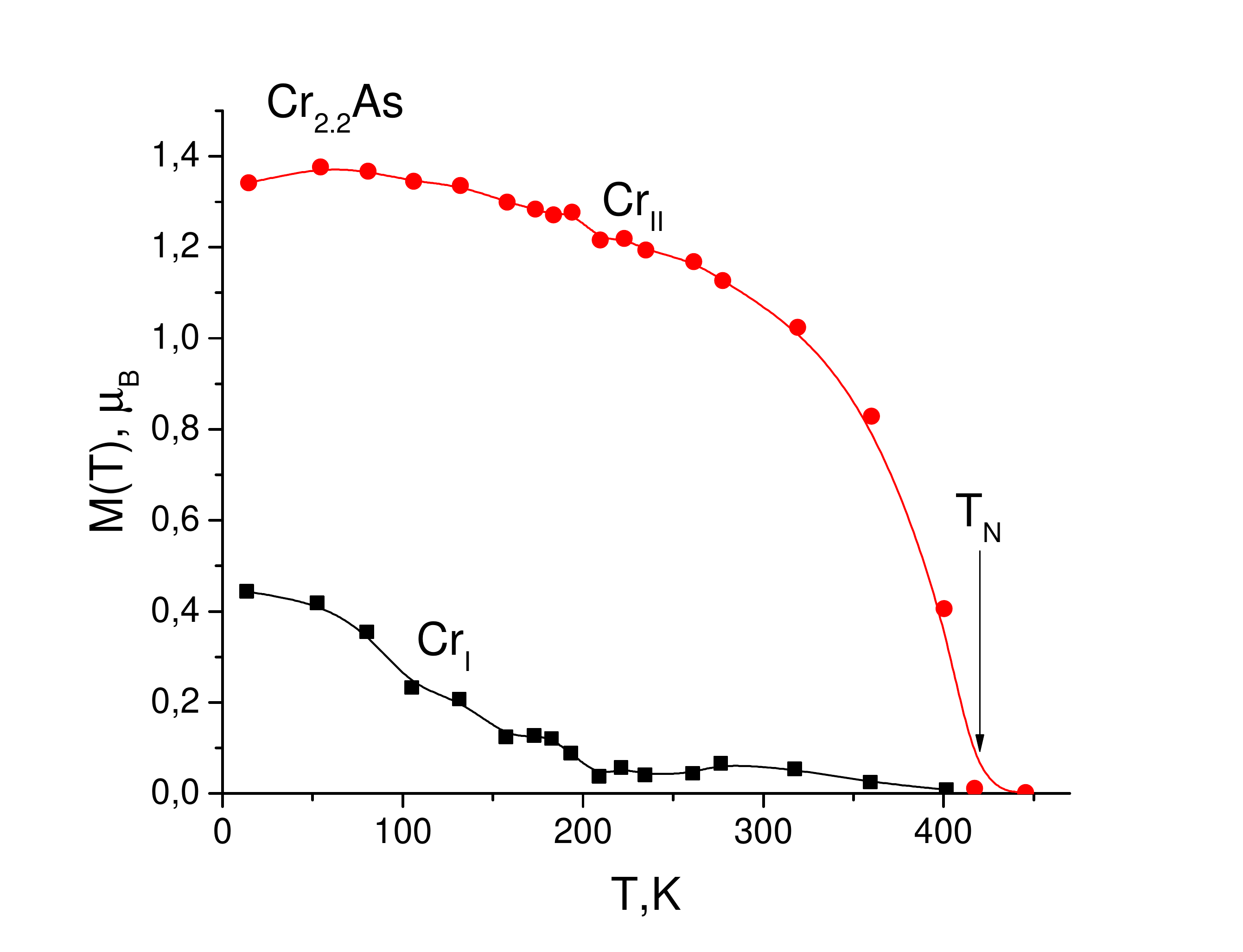}
\caption{\label{FIG:2}Temperature dependence of the sub-lattice magnetization of $Cr(\opencircle)$ and $Cr_{II}(\fullcircle)$ \cite{ishimoto1995anisotropic}.}
\end{figure}

Inter-metallic compounds of 3d metals with $As$ or $Sb$ possessing a tetragonal crystal structure of the $Cu_{2}Sb$-type (space symmetry group $D_{4h}^{7}$-P4/nmm) have garnered significant interest from the research community for their diverse magnetic structures. For example, $Mn_{2}Sb$ is a ferri-magnet \cite{darnell1963x}, and $Mn_{2}As$ \cite{austin1962magnetic} , $Fe_{2}As$ \cite{katsuraki1966magnetic} and $Cr_{2}As$ \cite{yamaguchi1972neutron} are anti-ferromagnets. 
The anti-ferromagnetic phase of $Cr_{2}As$ is distinguished by two aspects: the small magnetic moments (in terms of the Bohr magneton) of the atoms ($M(Cr_{I})$ = 0.4$μ_{B}$, $M (Cr_{II})$ = 1.34$μ_{B}$ [4]) and its magnetic structure Fig.\ref{FIG:1}(b). 
The peculiarity of the latter is that the effective molecular field between the $Cr_{I}$ and $Cr_{II}$ subsystems is compensated in the isotropic exchange approximation. 
This should lead to uncorrelated ordering of both subsystems, i.e. to the existence of two transition temperatures. 
\\ Initially, experimental studies revealed only one critical temperature ($T_{N}$ = 393K) \cite{yamaguchi1972neutron}, which led to the conclusion that the anisotropic exchange interaction of $Cr_{I}$-$Cr_{II}$ was significant \cite{valkov1992phenomenological}. 
However, later, in Ref.\cite{ishimoto1995anisotropic}, a second critical temperature was detected (Fig.\ref{FIG:2}, $T_{t}$ = 175K) corresponding to the ordering of the $Cr_{I}$ subsystem. 
Thus, the anisotropic part of the $Cr_{I}$-$Cr_{II}$ exchange interaction is not crucial and "works" only in the region of $T_{t} <T <T_{N}$, inducing a small magnetic moment of $Cr_{I}$ Fig.\ref{FIG:2}. 
Thus, $ab-initio$ calculation and subsequent analysis of the electronic structure and inter atomic exchange integrals in $Cr_{2}As$ are of interest to us.

\section{Crystal structure and Electronic structure of $Cr_{2}As$}

The basic parameters of the crystal and magnetic structures were taken from the experiment \cite{yamaguchi1972neutron,ishimoto1995anisotropic}. 
$Cr_{2}As$ has a $Cu_{2}Sb$-type tetragonal crystal structure with symmetry group $D_{4h}^{7}$ - $P4/nmm$, $a = 3.60$\AA, $c = 6.34$ \AA. 
$Cr_{I}$ occupies the positions of $2a (0,0,0)$ type, $Cr_{II}$ and $As$ occupy positions of $2c (0,0.5, z)$ type with parameters $z_{Cr}$ = 0.325, $z_{As}$ = 0.725, respectively. 
We considered all ten co-linear magnetic structures(Table \ref{tab:table1}), but the calculation was converged only for four of them(Table \ref{tab:table2}).
The magnetic moment of a specific atom at position $\mathbf{r}_0$ is given by \cite{valkov2012theoretical}:
$
M_{[0,1,0]}(\mathbf{r}_0) \sim \frac{\delta E (\mathbf{r}_0)}{J_0}\mu_B
$, 
where $\delta E = \int \limits^{E_F} d\epsilon \epsilon \left({LDOS_{\uparrow}\left({\mathbf{r}_0,\epsilon}\right)+LDOS_{\downarrow}\left({\mathbf{r}_0,\epsilon}\right)}\right)$ and $J_0$ is a single-center exchange integral discussed in details in the next section. 
Calculations of the electronic structure and exchange integrals in $Cr_{2}As$ are performed by fully relativistic Korringa-Kohn-Rostoker method (SPR-KKR software package \cite{ebert2012munich}). 
The atomic sphere approximation was used for the crystal potential. 
The exchange-correlation energy was calculated in the local density approximation without gradient corrections \cite{vosko1980accurate}.
\par
\begin{figure}[h!]
\centering
\includegraphics[width=1.0\textwidth]{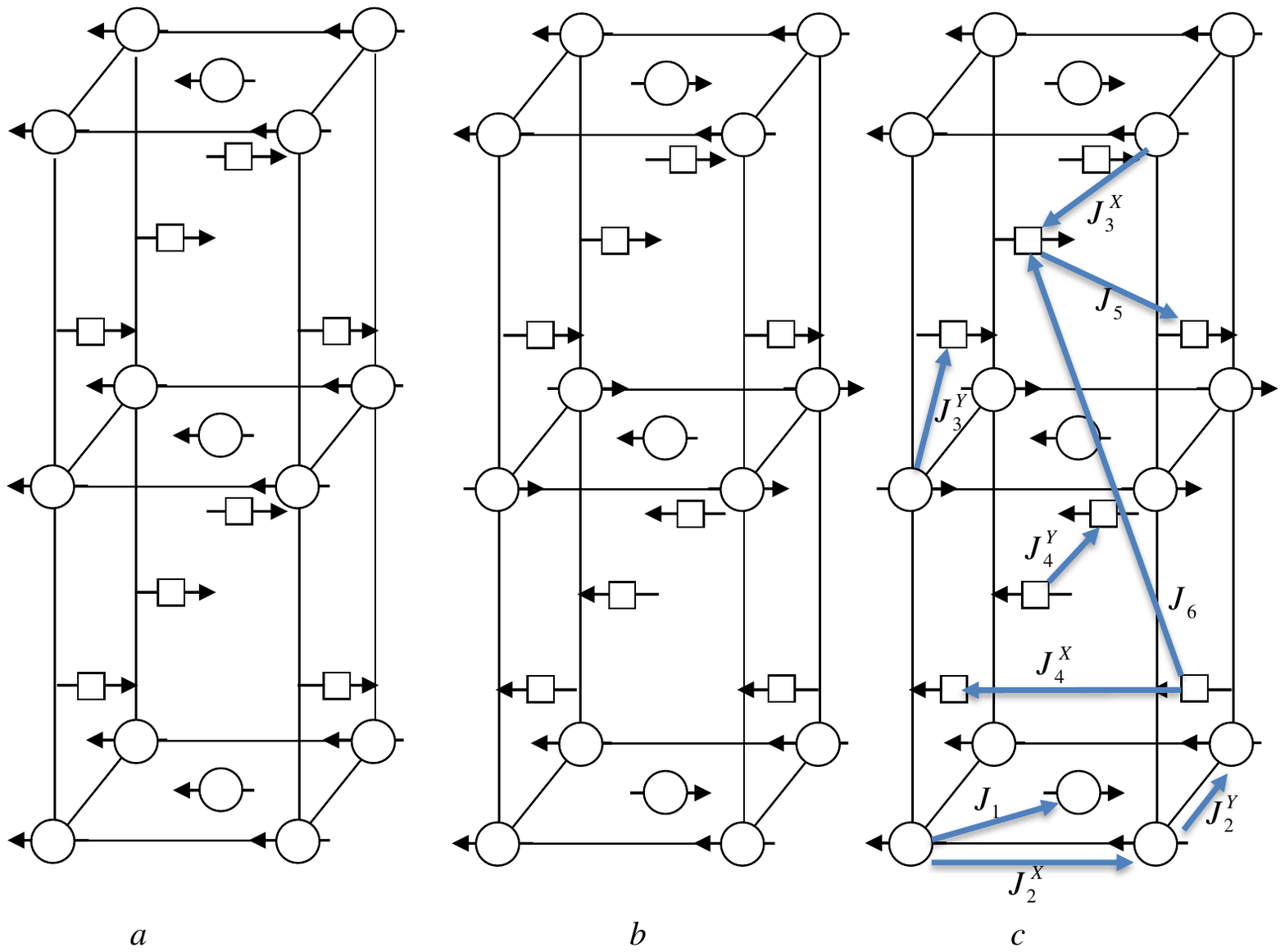}
\caption{\label{FIG:1}Magnetic structure of $Cr_{2}As$: FIM(a) and AF(b). $\opencircle$ and $\opensquare$ indicate $Cr_{I}$ and $Cr_{II}$, respectively. Only metal atoms are shown. Main inter-atomic exchange interactions is shown on figure(c).}
\end{figure}

\begin{figure}[h!]
\centering
\includegraphics[width=0.5\textwidth]{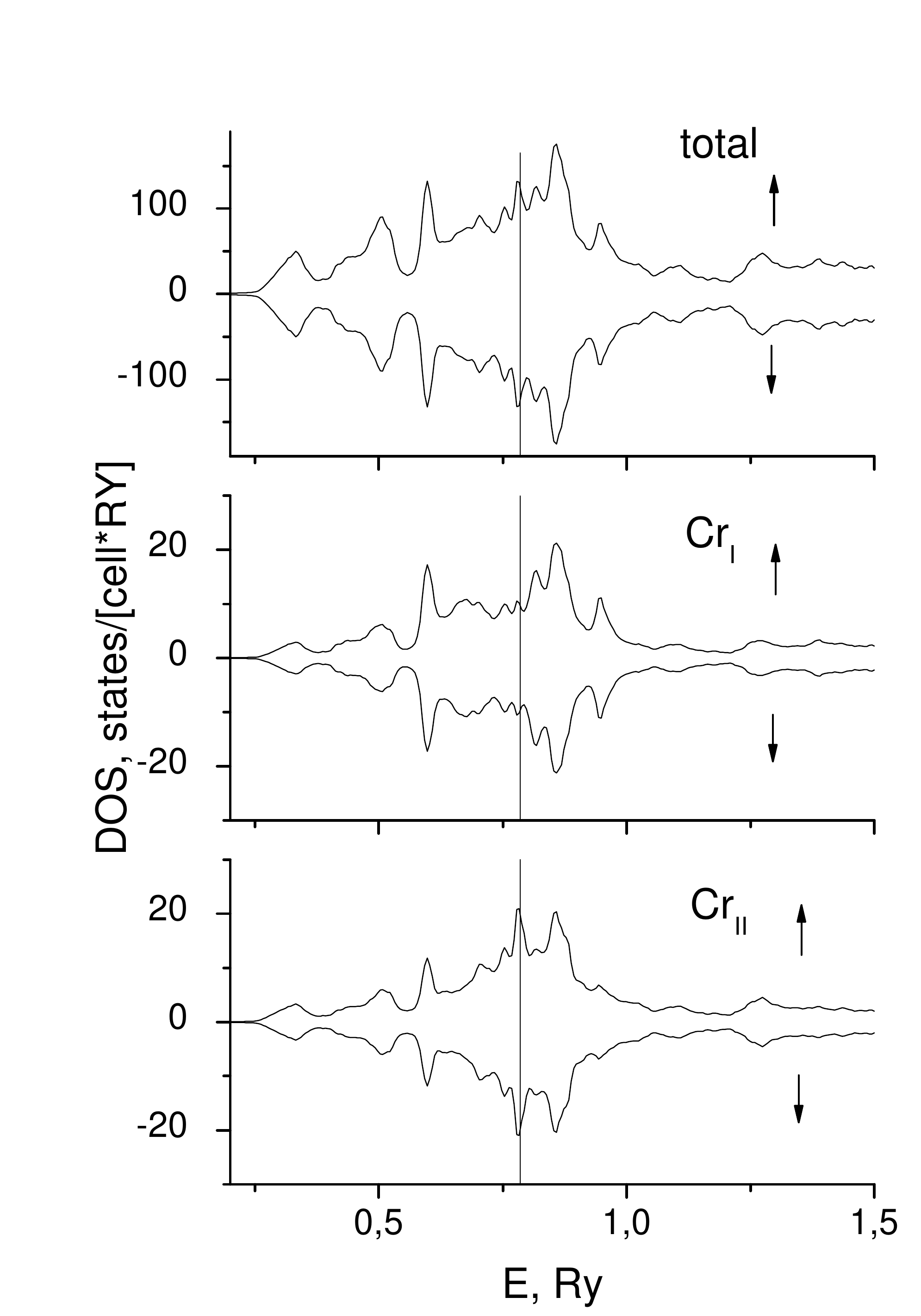}
\caption{\label{FIG:3}Density of electronic states for nonmagnetic $Cr_{2}As$. Vertical line indicates Fermi energy ($E_F$), arrows point to the corresponding spin orientation.}
\end{figure}

\begin{figure}
\centering
\subfloat[]
{
\includegraphics[width=0.45\textwidth]{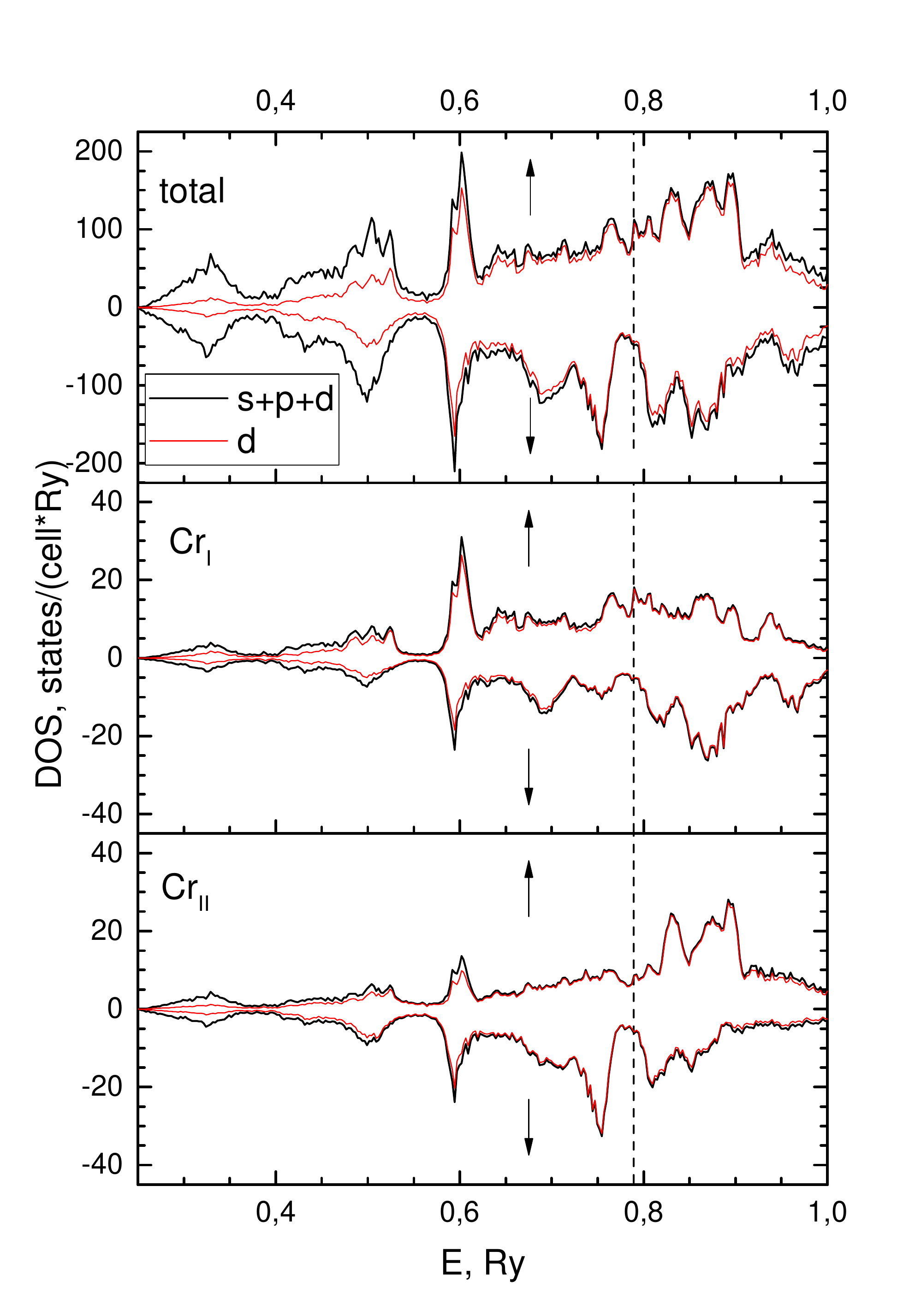}
}
\subfloat[]
{
\includegraphics[width=0.45\textwidth]{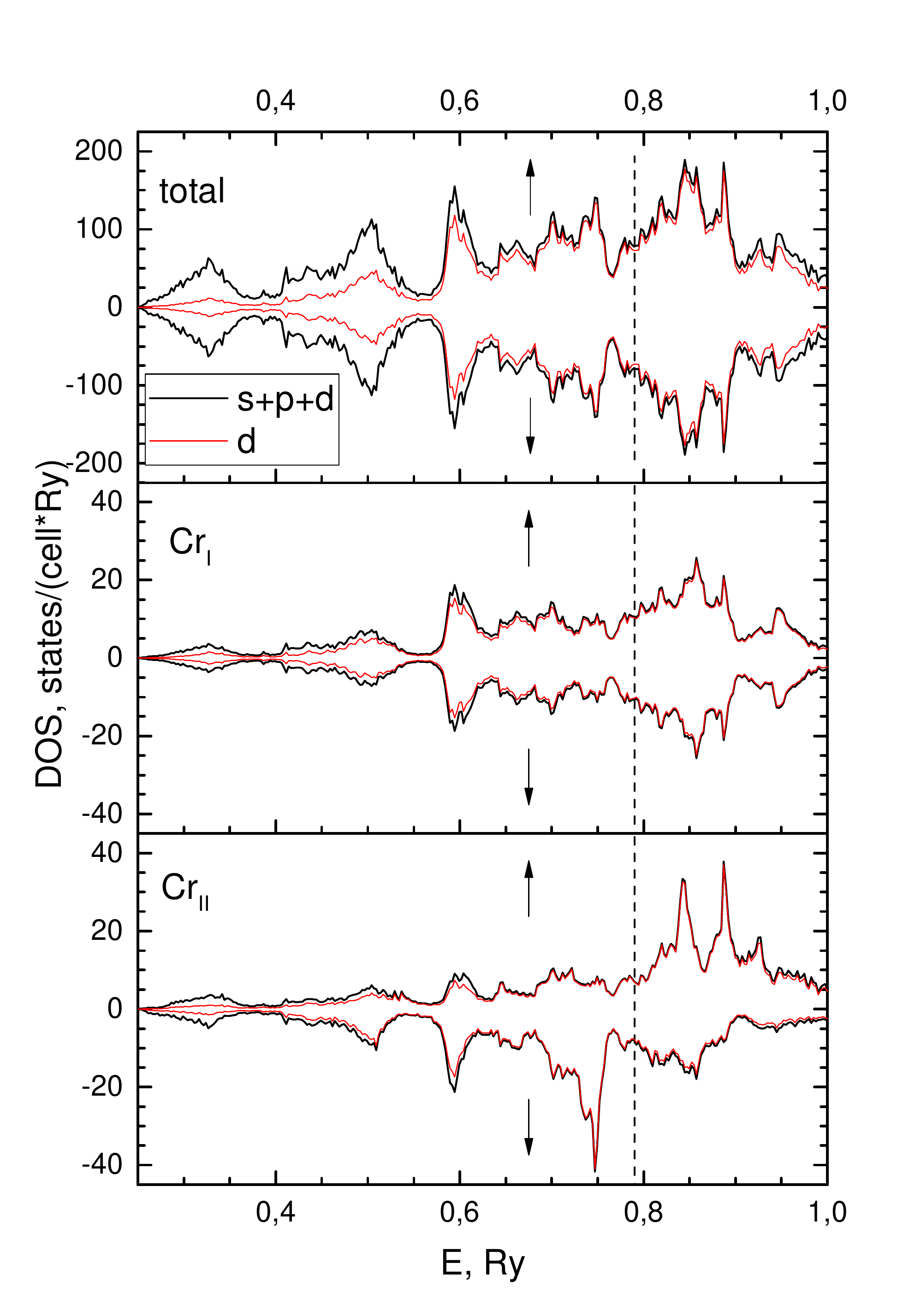}
}
\\
\subfloat[]
{
\includegraphics[width=0.45\textwidth]{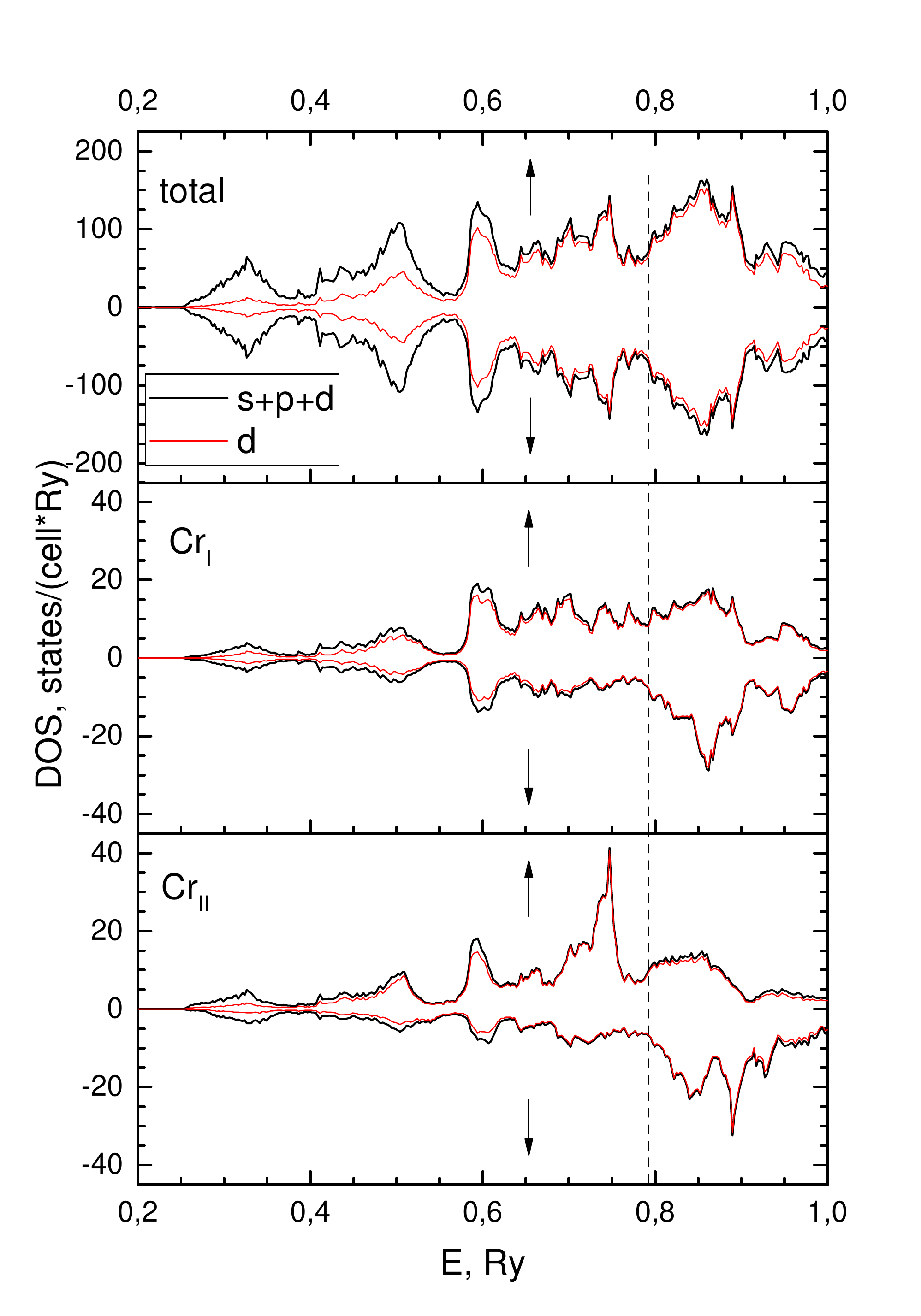}
}
\subfloat[]
{
\includegraphics[width=0.45\textwidth]{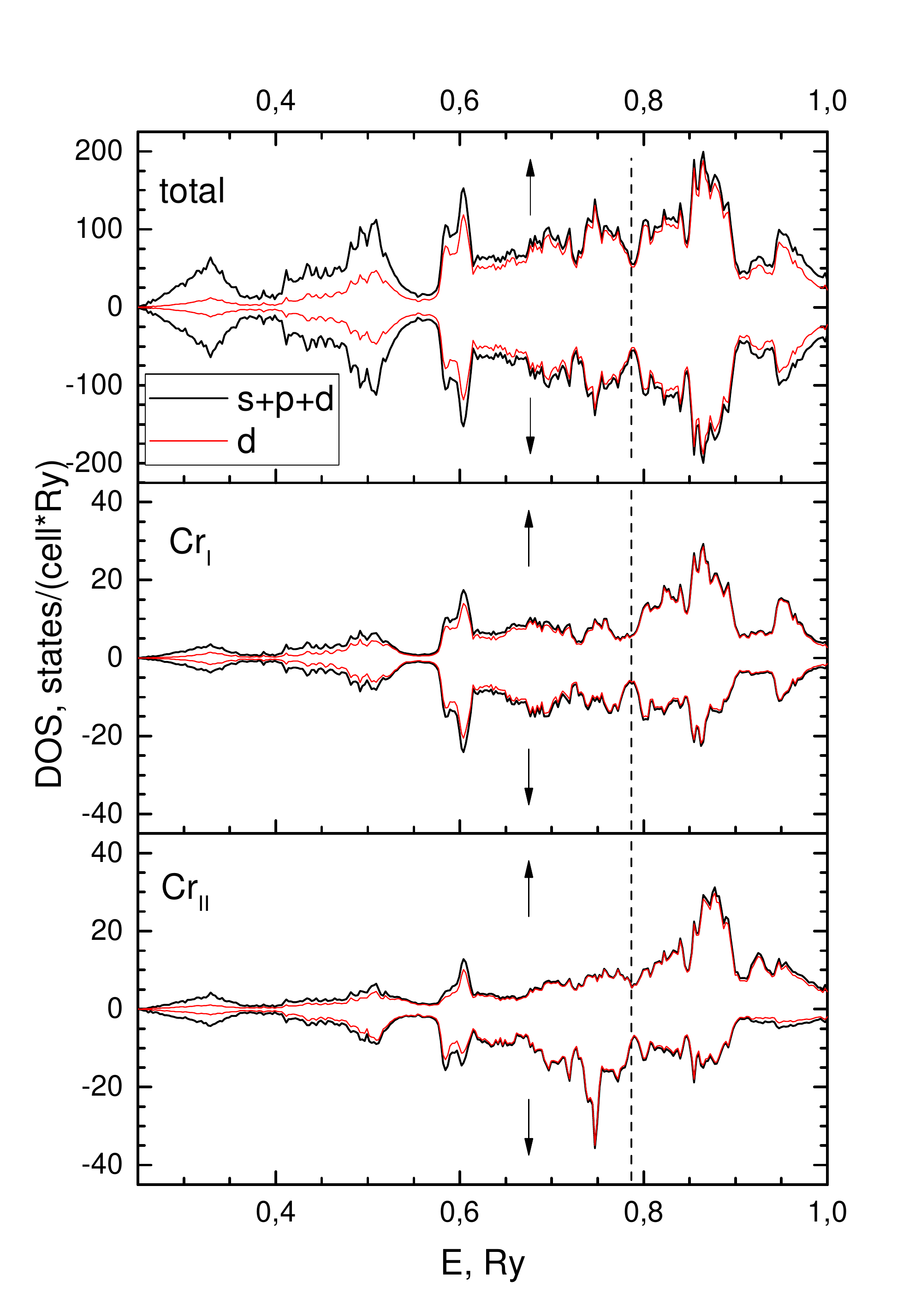}
}
\caption{\label{FIG:4}  Density of electronic states for ferrimagnetic(a) and anti-ferromagnetic: AF1(b), AF3(c), AF4(d) $Cr_{2}As$. Notation follows that of Fig.\ref{FIG:3}}
\end{figure}

 
According to the results of calculations, the lowest energy has an anti ferromagnetic structure AF3(Table \ref{tab:table2}, Fig.\ref{FIG:1}(b)), which agrees with the experimental data. 
At the same time, the calculation shown the instability of the ferromagnetic phase. 
Therefore, as a starting point, a ferrimagnetic structure of the $Mn_{2}Sb$-type was considered (Fig.\ref{FIG:1}(a)). \\
The electronic structure of nonmagnetic $Cr_{2}As$ is shown in Fig.\ref{FIG:3} and its magnetic counterpart is depicted in Fig.\ref{FIG:4}. 
The conduction band is located above $0.25 Ry$ and is formed mainly by the $3d$-states of $Cr$ and $4p$-states of $As$, that indicates strong $p-d$-hybridization in this compound. 
In general, the electronic structure is typical for pnictides of transition metals and agrees with the results of other authors \cite{shirai1993electronic}. 
The $Cr_{2}As$ compound is a metal and all features of the magnetism of transition metal compounds are inherent in it. Chromium d-electrons create magnetic and transport properties. 
They are de-localized and so are the magnetic moments of chromium atoms (Table  \ref{tab:table2}).\\
The magnetic moments of chromium in the ferrimagnetic ($M(Cr_{I})$ = 0.709$\mu_{B}$, $M(Cr_{II})$ = 1.438$\mu_{B}$) and anti ferromagnetic (AF3, $M(Cr_{I})$ = 0.937$\mu_{B}$, $M(Cr_{II})$ = 1.633$\mu_{B}$) phases agree with the experimental data \cite{yamaguchi1972neutron} ($M(Cr_{I})$ = 0.4$\mu_{B}$, $M(Cr_{II})$ = 1.34$\mu_{B}$) and the results of other calculations by LAPW[10] ($M(Cr_{I})$ = 0.33$\mu_{B}$, $M(Cr_{II})$ = 1.37$\mu_{B}$) and KKR[11]($M(Cr_{I})$ = 0.43$\mu_{B}$, $M(Cr_{II})$ = 1.75$\mu_{B}$) methods. 
The high value of the $Cr_{I}$ magnetic moment is related to the used approximation for potential of crystal lattice \cite{yildirim2009frustrated}.\\

\section{Inter-atomic Exchange Interactions and critical magnetization temperature}

\begin{figure}
\centering
\subfloat[]
{
\includegraphics[width=0.5\textwidth]{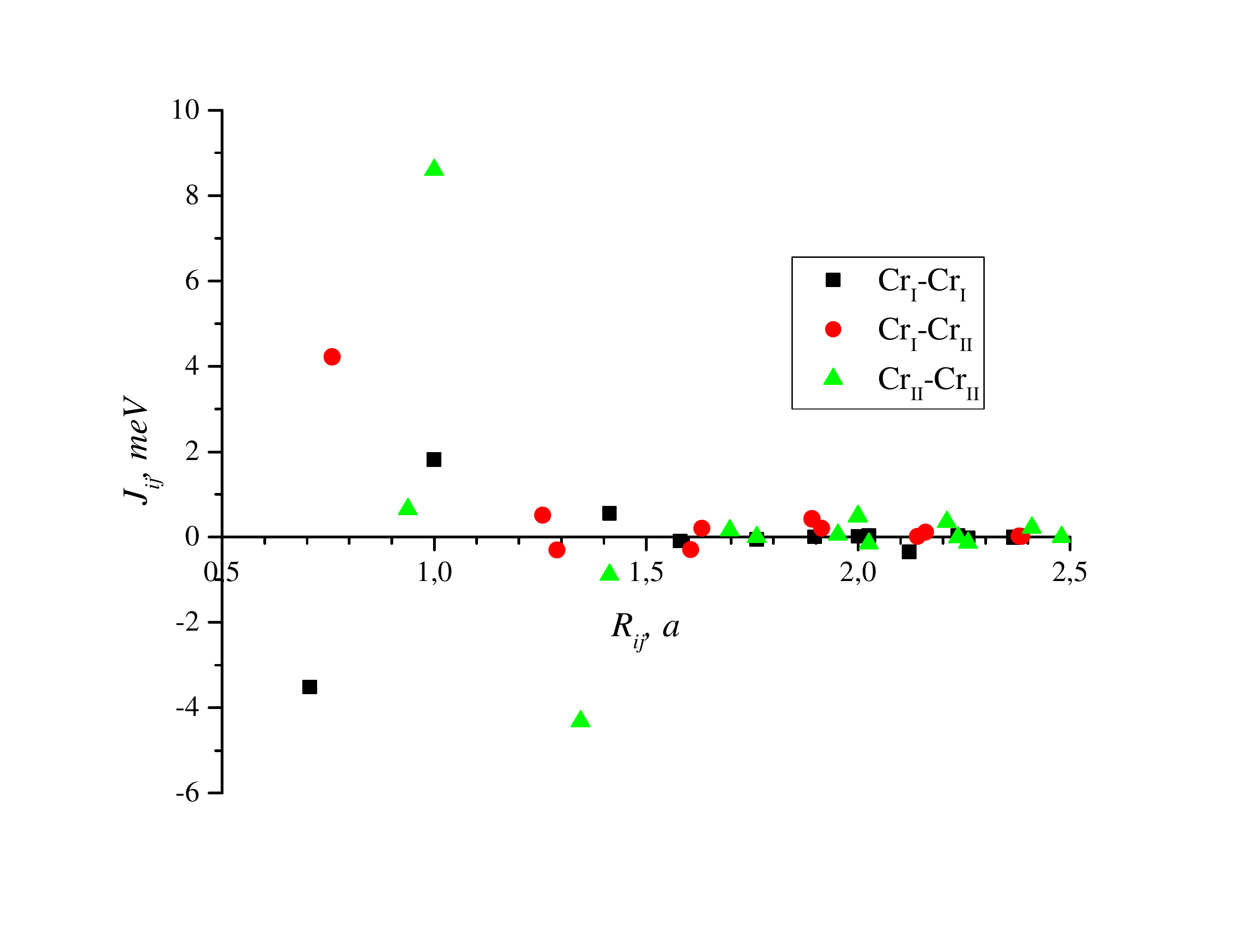}
}
\subfloat[]
{
\includegraphics[width=0.5\textwidth]{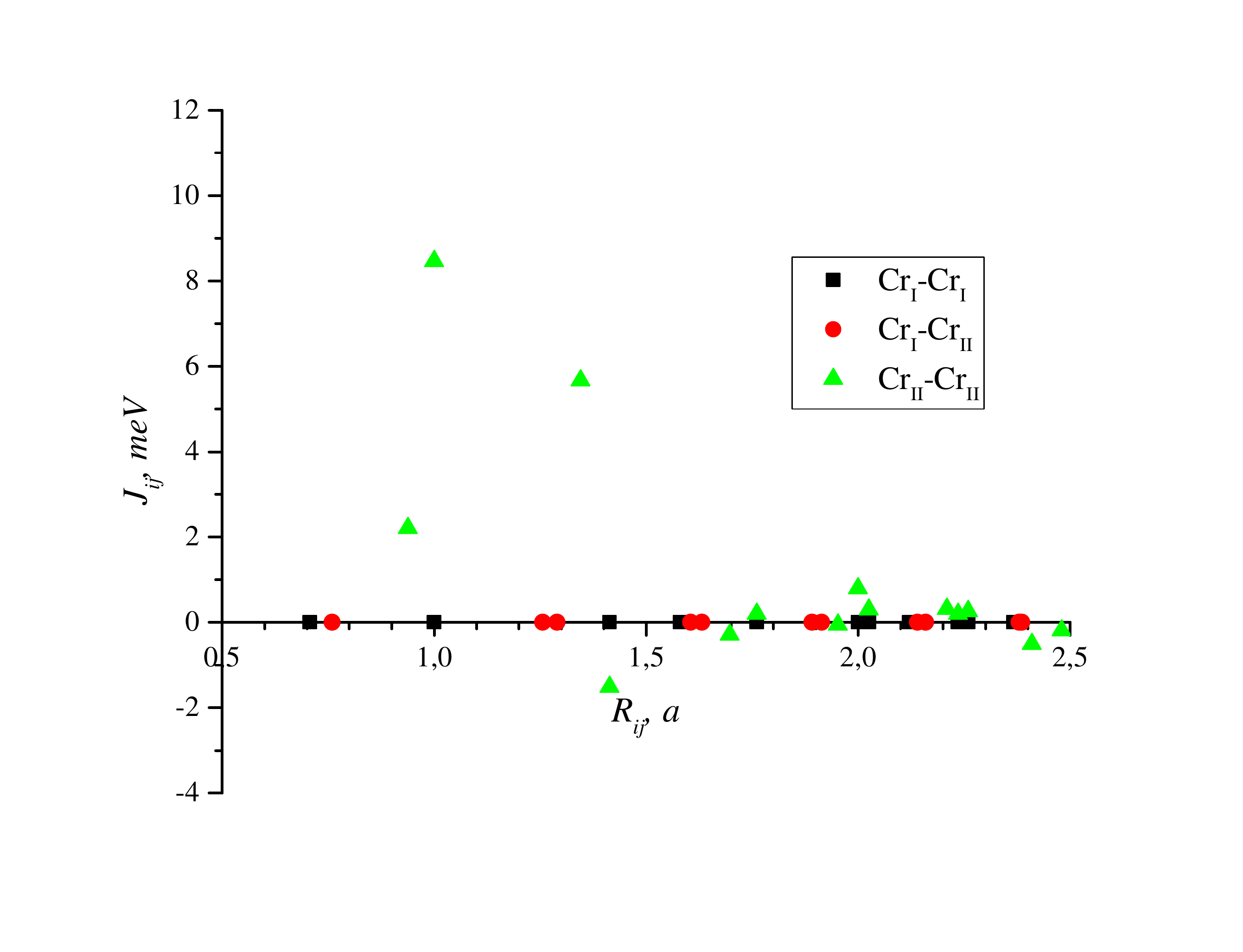}
}
\\
\subfloat[]
{
\includegraphics[width=0.5\textwidth]{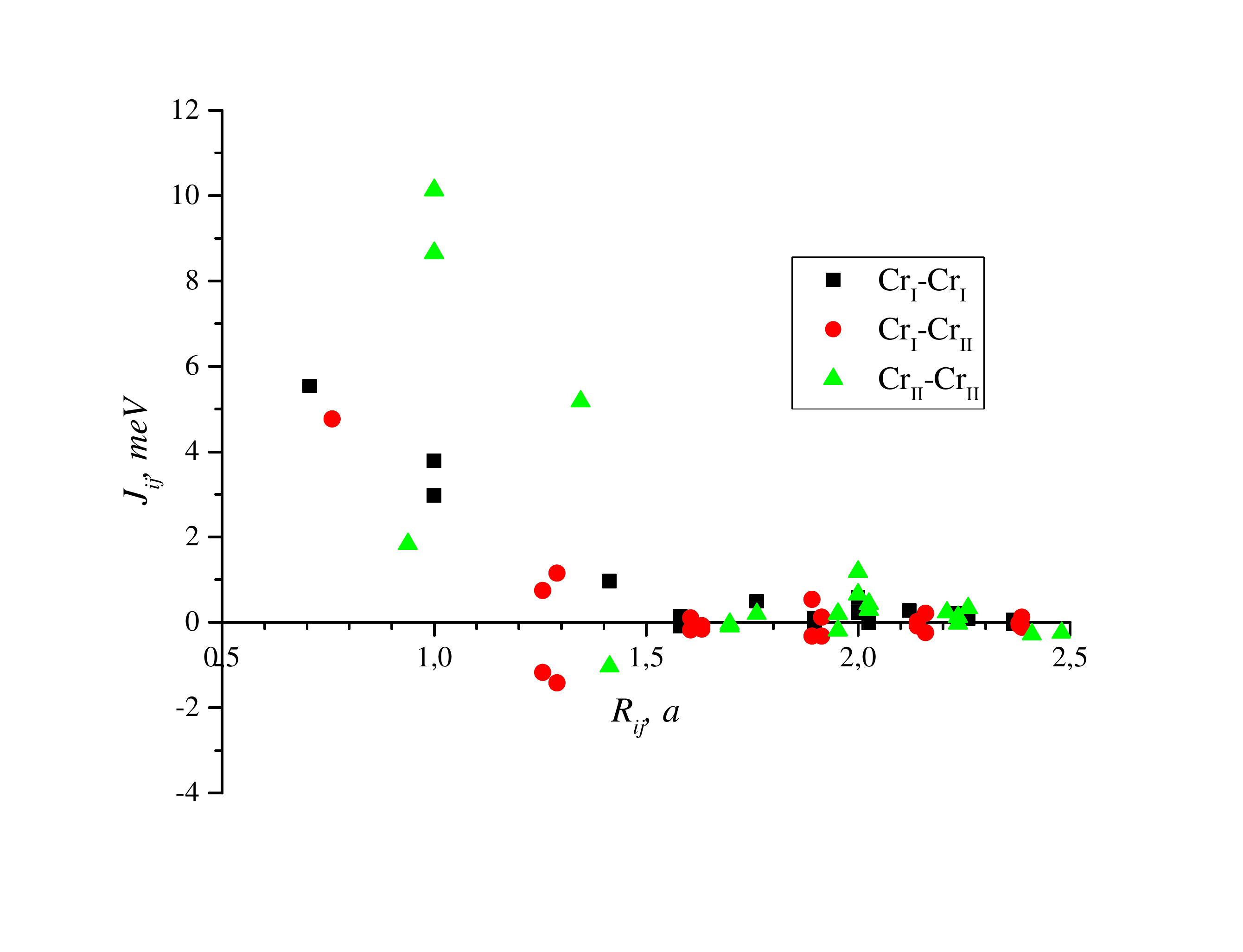}
}
\subfloat[]
{
\includegraphics[width=0.5\textwidth]{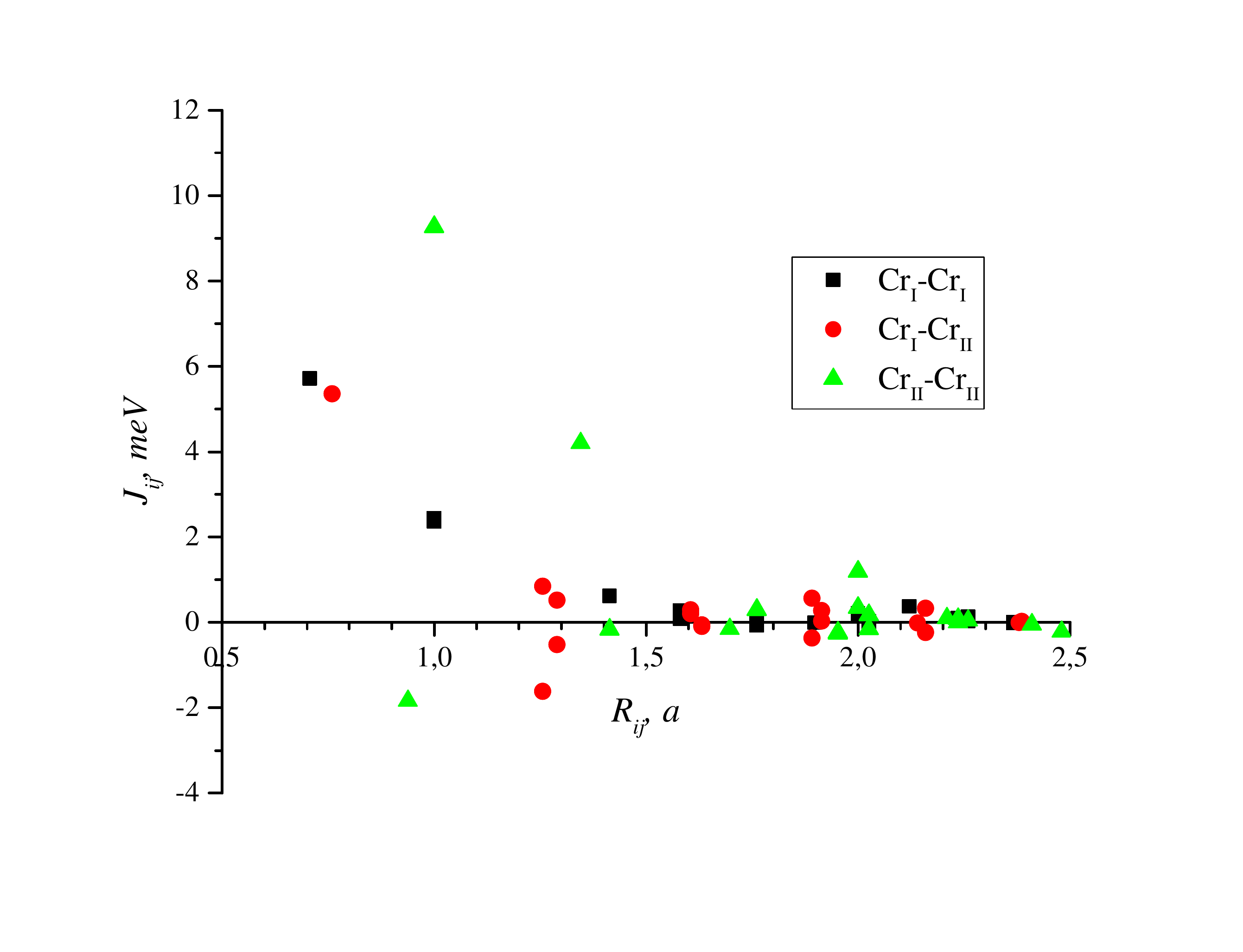}
}
\caption{\label{FIG:5}  Dependence of inter-atomic exchange interactions in $Cr_{2}As$ from inter-atomic distance (in lattice units a) in ferrimagnetic(a) and anti-ferromagnetic: AF1(b), AF3(c), AF4(d) structures.}
\end{figure}

\begin{figure}
\centering
\subfloat[]
{
\includegraphics[width=0.5\textwidth]{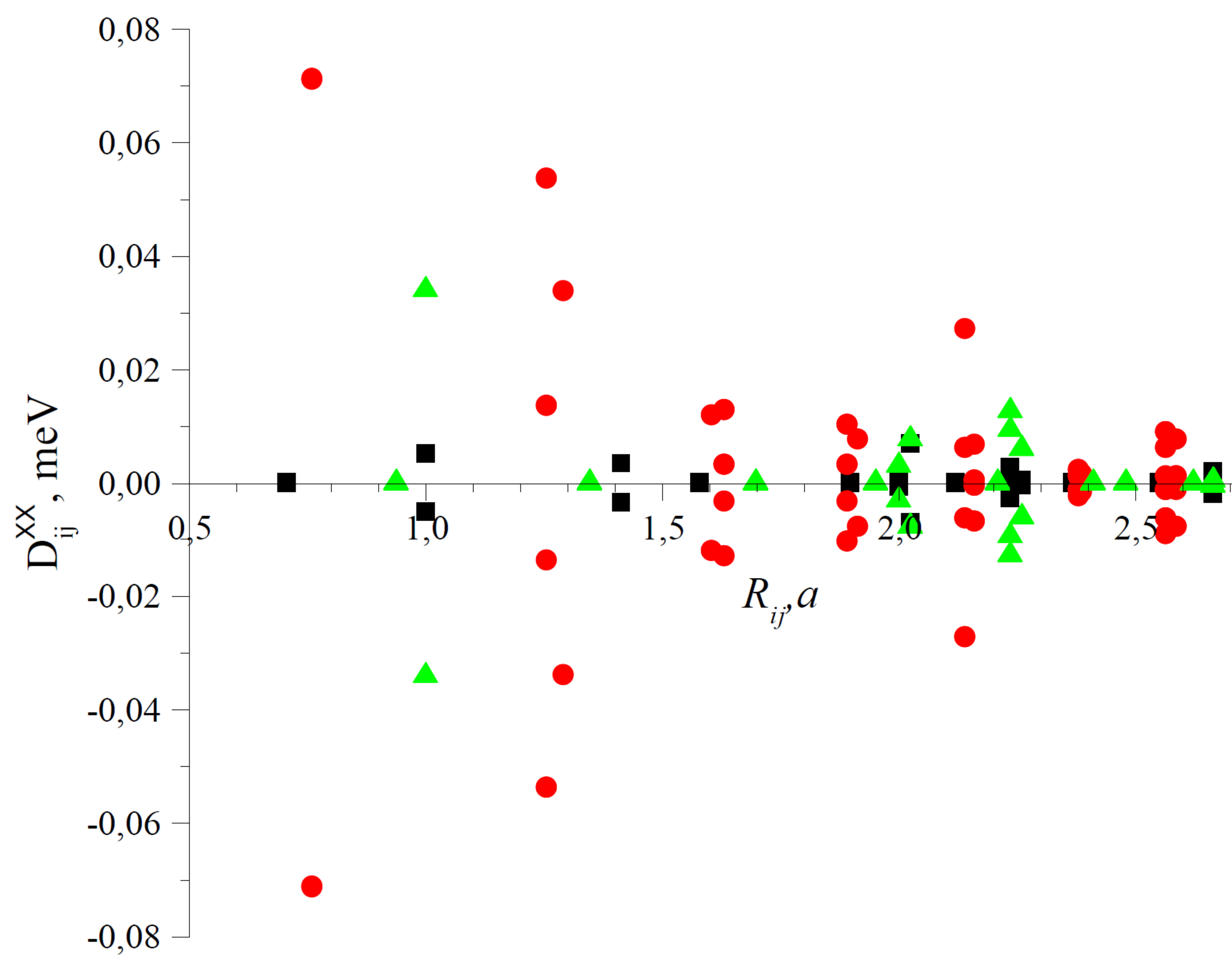}
}
\\
\subfloat[]
{
\includegraphics[width=0.5\textwidth]{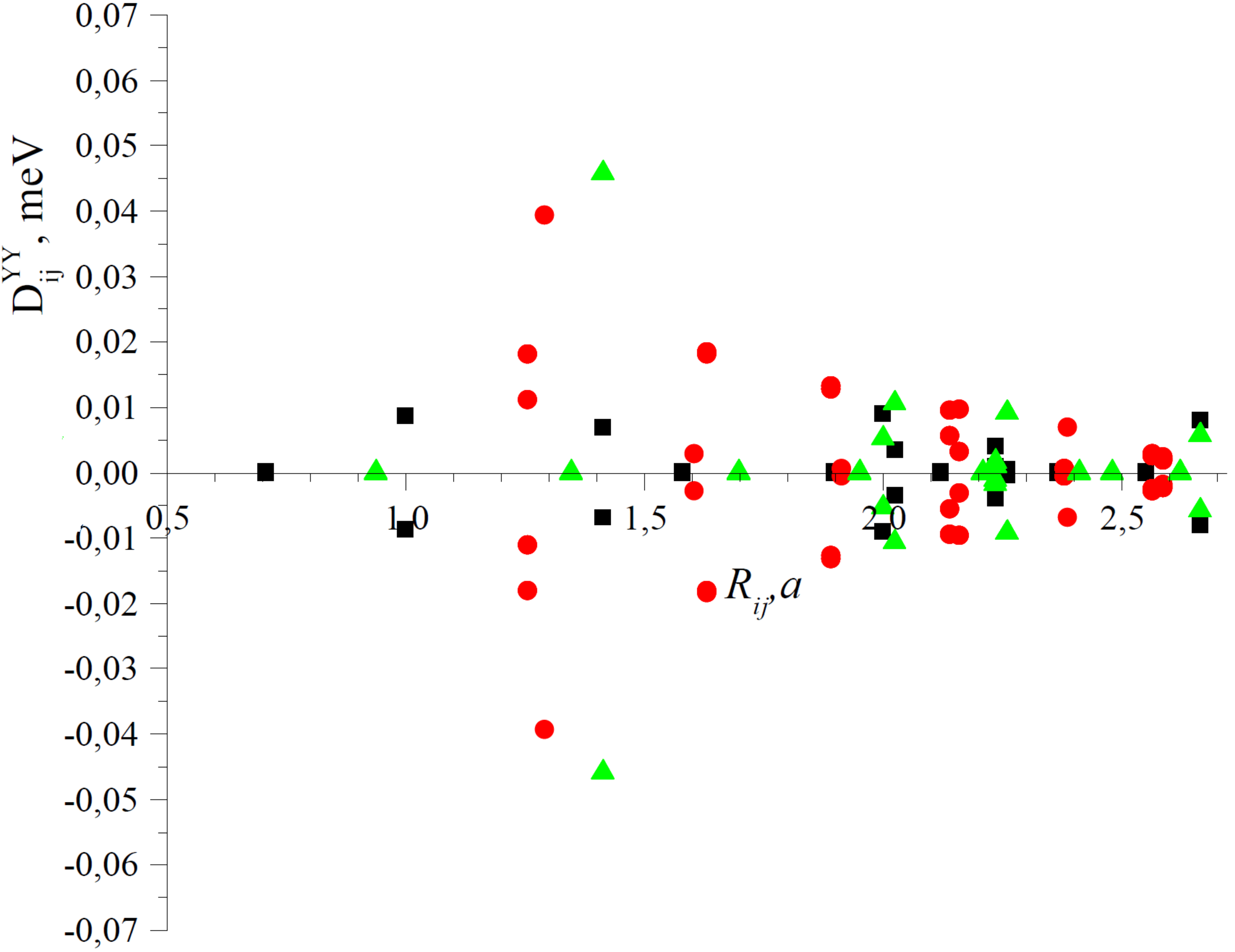}
}
\\
\subfloat[]
{
\includegraphics[width=0.5\textwidth]{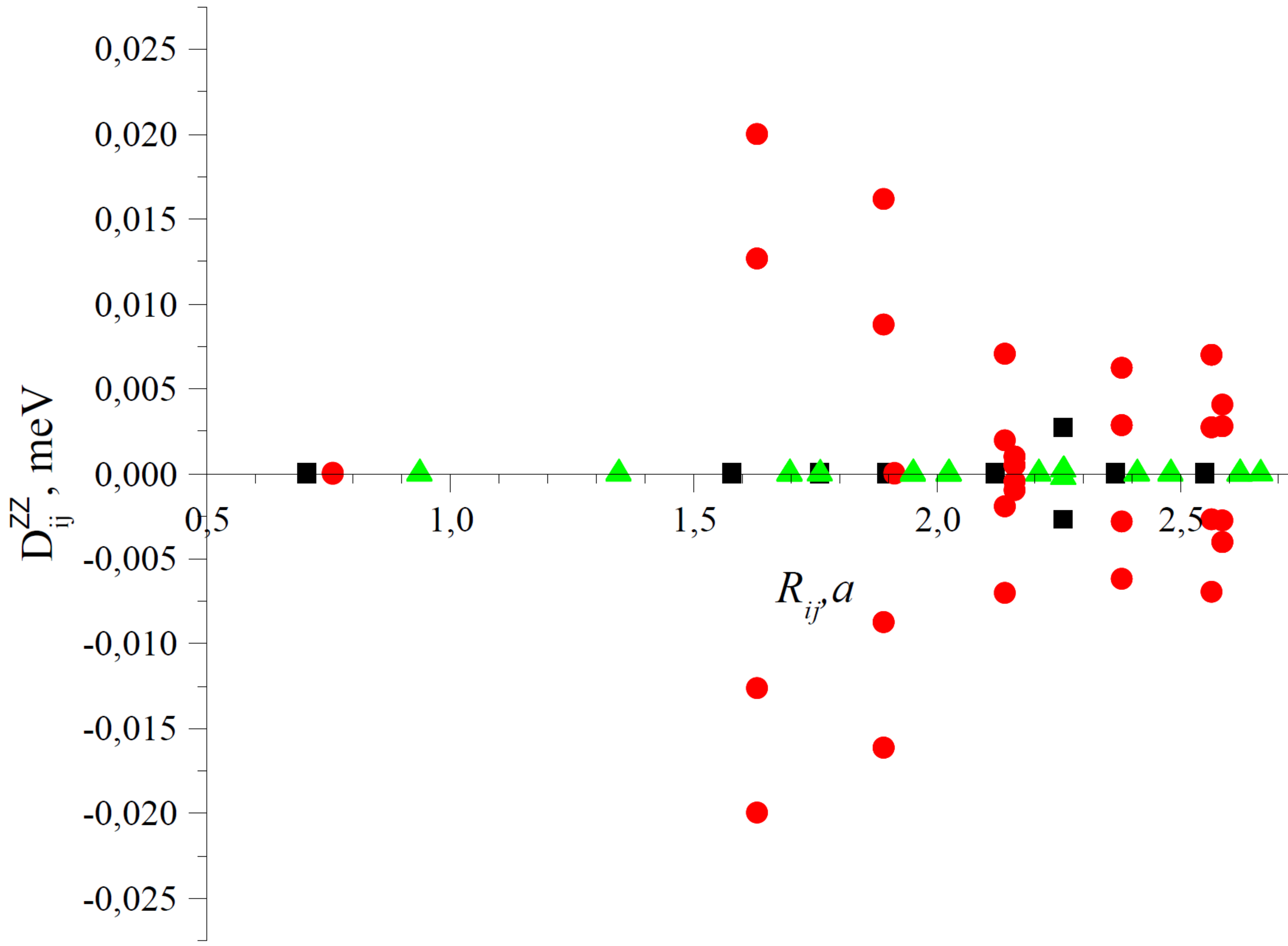}
}
\caption{\label{FIG:6}  Dependence of diagonal component of Dzyaloshinskii-Moriya interaction  $D_{ij}^{X}(a)$,  $D_{ij}^{Y}(b)$, $D_{ij}^{Z}(c)$ in AF3-structure of $Cr_{2}As$ from interatomic distance (in lattice units a)}
\end{figure}

The inter-atomic exchange integrals were calculated by the method \cite{liechtenstein1987local}, based on the calculation of the second derivative of the total energy functional from the deviations of the selected pair of spins from equilibrium. 
The effective classical Heisenberg Hamiltonian for metals/alloys is:
\begin{equation}
H=-\frac{1}{2}\sum_{i\ne j} \left[{ (J_{ij}^{iso}+J_{ij}^{sym}) \mathbf{e}_{i} \cdot \mathbf {e}_{j} + D_{ij} \mathbf{e}_{i}\times\mathbf{e}_{j}}\right], 
\end{equation}
, where the direction of local magnetic moments on $i$ site is described by unit vectors $\mathbf{e}_{i}$. 
The Hamiltonian contains isotropic ($J_{ij}^{iso}$), symmetric anisotropic ($J_{ij}^{sym}$) exchange coupling as well as Dzyaloshinskii-Moriya interaction ($D_{ij}$). 
The Dzyaloshinskii-Moriya interaction $D_{ij} \sim \sum \limits_{\beta \ne 0} J_{0\beta} (\mathbf{r}_{j\beta} \cdot \mathbf{r}_{\beta i})$ is reported in Fig.\ref{FIG:6}. For brevity those are depicted for AF3 configuration.
The dependence of the exchange couplings from inter-atomic distance is shown in Fig.\ref{FIG:5} for the FIM and AF-states.
Now let us fix the position of an atom of interest as $i=0$ and consider exchange parameter with nearby atoms $J_{0j}$ (for convenience we may omit zero subscript when it causes no confusion).
In ferrimagnetic phase the exchange interaction that ensures the bonding of $Cr_{I}$ and $Cr_{II}$ sub-lattices is $J_{3}=4.22$ meV in the first coordination sphere and rapidly decreases in subsequent (Fig.\ref{FIG:5}). 
The negative exchange interaction between the nearest $Cr_{I}$ atoms ($J_{1}=-3.52$ meV) ensures their "unusual" orientations in the AF structure (in a nonmagnetic crystal, the $Cr_{I}$ atoms are symmetrically identical). 
\\
Among the exchange integrals that are important for describing the magnetic structure of $Cr_{2}As$ (Figure\ref{FIG:5}, Table 3), only $J_{6}$ can be considered to be created by the double indirect exchange of $Cr_{II} - As - Cr_{II}$ atoms. 
For the other significant exchange integrals, the distance between the cations is smaller or comparable with the distances of the cation-anion \cite{sato2010first}. 
Therefore, it is more appropriate to talk about the mechanism of classical d-d exchange, which can be either ferromagnetic or anti-ferromagnetic (ferromagnetic iron and anti-ferromagnetic manganese or chromium).
In other words the very existence of FM and AFM only phases (lack of intermediate magnetic structure) is due to large  value of of the $d-$band electronic filling.
This differs substantially from the case of diluted magnetic semiconductors such as $Ga_{1-x}Mn_{x}As$. In the latter case the interaction of magnetic atoms via nonmagnetic As becomes the dominant mechanism due to low concentration of manganese atoms \cite{sato2010first}. 
\\
The main topic of interest in $Cr_{2}As$ is the interrelation between two chromium sub-lattices, which can be provided only by anisotropic exchange interaction \cite{valkov1992phenomenological,ishimoto1995anisotropic}. 
To estimate its value, the interatomic exchange integrals for the different AF-states (Fig.\ref{FIG:5}, Table \ref{tab:table3}) was calculated. 
In the case of AF3 structure, for an arbitrary $Cr_{I}$, in one half of the nearest $Cr_{II}$ atoms, the magnetic moments are parallel, while for the second one they are anti parallel to the magnetic moment of the chosen atom. 
The observed difference in the exchange integrals along X and Y axes $J_{3}^{X}(Cr_{I}-Cr_{II}) =4.77$ meV and $J_{3}^{Y}(Cr_{I}-Cr_{II}) =-6.36$ meV confirms the existence of anisotropy of the $Cr_{I}$-$Cr_{II}$ exchange interaction \cite{ishimoto1995anisotropic}. 
However, its magnitude is insufficient to ensure simultaneous transition of sub-lattices to a magnetically ordered state (Fig.\ref{FIG:2}). 
The anisotropy part of $Cr_{II}-Cr_{II}$ exchange interaction much smaller $J_{4}^{X}(Cr_{II}-Cr_{II}) =10.13$ meV, $J_{4}^{Y}(Cr_{II}-Cr_{II}) =8.66$ meV. \\ 
We estimate the temperatures of the magnetic ordering of the chromium sub-lattices by the well established expression formula:
$
T_{0}=\frac{2}{3}J_{0}, 
$
which is valid for the Heisenberg model with classical spins \cite{liechtenstein1987local}. 
Here $J_{0}=\sum\limits_{j \neq 0} J_{0j}$ is the effective exchange interaction of the chosen atom with the entire crystal. 
The effective ordering temperatures are $T(Cr_{I})$ = 180K and $T(Cr_{II})$ = 382K versus experimentally observed $T(Cr_{I})$ = 175K and $T(Cr_{II})$ = 393K.
The detailed analyses of orientation dependence of magnetic moments and interatomic exchange interaction(Table \ref{tab:table2}, Table \ref{tab:table3}) shown that magnetic moment of $Cr_{I}$ is induced by exchange interaction within $Cr_{II}$ sub-lattice. 
\par 
As was indicated in Ref. \cite{sato2004low} estimate of the critical temperature on the basis of mean-field approximation is inaccurate and gives too high or too low value, depending on the model used.
However for $Cr_{2}As$ estimated (experimental results as shown in Fig.\ref{FIG:2}) critical temperature for $Cr_{I}$ and $Cr_{II}$ differ by no more than 3\%. 
Such accuracy justifies the mean field approximations used for calculation of the exchange integrals as compared to more involved Monte Carlo simulations.
The Monte Carlo method improves convergence provided the underlying code considers:  a) the changes in the electronic structure occurring when the magnetic moments of chromium atoms are reoriented; 
b) the corresponding changes in exchange integrals; c) the effects of percolation in our case are not as important as in diluted magnetic semiconductors\cite{sato2004low}.
\par
SPR-KKR calculation is based on the LDA.
Self-interaction corrections for Kohn-Sham density functional theory were reported in literature\cite{toyoda2006curie,an2006exchange}. Their physical meanings, formulations, and applications to the critical temperature of the relevant phase transition has been discussed. In essence, the self-interaction corrections get rid of the self-interaction error, which is the sum of the Coulomb and exchange self-interactions that remains because of the use of an approximate exchange functional. The most frequently used self-interaction correction is the Perdew-Zunger correction. However, this correction leads to instabilities in the electronic state calculations of molecules. To avoid these instabilities, the authors provided several self-interaction corrections on the basis of the characteristic behaviors of self-interacting electrons, which have no two-electron interactions. These include the von Weizsäcker kinetic energy and long-range (far-from-nucleus) asymptotic correction. Applications of self-interaction corrections have shown that the self-interaction error has a serious effect on the states of core electrons, but it has a smaller than expected effect on valence electrons as is the case described in our manuscript. The distribution of self-interacting electrons indicates that they are near atomic nuclei rather than in chemical bonds and has limited effect on the mean field theory. We believe that at this point we shall leave those effects to further discussions. 

\section{Conclusion}
In the literature, there is a paper \cite{fruchart2005magnetic} devoted to theoretical analysis by the Berto method \cite{kallel1974helimagnetism} of possible magnetic structures and the conditions for their realization in magnets with a tetragonal lattice of the $Cu_{2}Sb$-type. 
The Ref. \cite{fruchart2005magnetic} indicates as the crucial parameter for the existence of the AF structure the large value of the indirect exchange of $Cr_{II}-As-Cr_{II}$($J_6$ in our notation). 
However, the applicability of their  conclusions to the description of the AF structure in $Cr_{2}As$ seems doubtful due to the isotropic exchange approximation used \cite{fruchart2005magnetic} to obtain them. 
Preliminary analysis of magnetic structures by the Berto method \cite{kallel1974helimagnetism} showed that for the realization of the experimentally observed AF structure in $Cr_{2}As$, the existence of anisotropy of the exchange interaction ($J^{\uparrow\uparrow}(Cr_{I}-Cr_{II}) \neq J^{\uparrow\downarrow}(Cr_{I}-Cr_{II})$) is a necessary condition.

\section{ Data Availability}The data used to support the findings of this study are included within this article. It was obtained using open source software 'The munich spr-kkr package, version 7.7'.

\section{Conflicts of Interest} The authors declare that there is no conflict of interest regarding the publication of this paper. 

\section{Acknowledgments}
This work was funded by PSC-CUNY grant and Fordham Research Grant (FRG).

%


\newpage 
\section{Tables}

\begin{table}[ht!]
	\begin{center}
		\caption{Possible collinear magnetic structures}\label{tab:table1}				
		\begin{tabular}{c|c|c|c|c|c|c|c|c|c|c|c|c} 
			N &	Atom & Coordinates & FM & FIM & AF1 & AF2 & AF3 & AF4 & AF5 & AF6 & AF7 & AF8 \\
			\hline
			1 & $Cr_{I}$  & (0,0,0)       & + & - & + & + & + & + & + & + & + & + \\
			2 & $Cr_{I}$  & (0.5,0.5,0)   & + & - & + & + & - & - & - & + & + & - \\
			3 & $Cr_{II}$ & (0,0.5,0.325) & + & + & - & + & + & - & + & + & - & + \\
			4 & $Cr_{II}$ & (0.5,0,0.675) & + & + & - & + & + & + & - & - & + & + \\ 
			5 & $Cr_{I}$  & (0,0,1.0)     & + & - & - & - & - & - & - & - & - & - \\
			6 & $Cr_{I}$  & (0.5,0.5,1.0) & + & - & - & - & + & + & + & - & - & + \\
			7 & $Cr_{II}$ & (0,0.5,1.675) & + & + & + & - & - & + & - & - & + & - \\
			8 & $Cr_{II}$ & (0.5,0,1.325) & + & + & + & - & - & - & + & + & - & - \\ 
		\end{tabular}
	\end{center}
\end{table}

\begin{table}[ht!]
\setlength\tabcolsep{2pt} 	
\footnotesize  
	\centering
	\caption{Calculated parameters of magnetic structures in $Cr_{2}As$}\label{tab:table2}			
		\begin{tabular}{c|c|c|c|c|c|c} 
			N &	Atom & Coordinates & FIM & AF1 & AF3 &  AF4 \\
			\hline
			1 & $M(Cr_{I}),\mu_{B}$  & (0,0,0)       & -0.709 & 0.0   & 0.937 & -0.904  \\
			2 & $M(Cr_{I}),\mu_{B}$  & (0.5,0.5,0)   & -0.709 & 0.0   &-0.937 &  0.904  \\
			3 & $M(Cr_{II}),\mu_{B}$ & (0,0.5,0.325) & 1.438  &-1.598 & 1.633 & -1.653  \\
			4 & $M(Cr_{II}),\mu_{B}$ & (0.5,0,0.675) & 1.438  &-1.598 & 1.633 &  1.653  \\
			5 & $M(Cr_{I}),\mu_{B}$  & (0,0,1.0)     & -0.709 & 0.0   &-0.937 &  0.904  \\
			6 & $M(Cr_{I}),\mu_{B}$  & (0.5,0.5,1.0) & -0.709 & 0.0   & 0.937 & -0.904  \\
			7 & $M(Cr_{II}),\mu_{B}$ & (0,0.5,1.675) & 1.438  & 1.598 &-1.633 &  1.653  \\
			8 & $M(Cr_{II}),\mu_{B}$ & (0.5,0,1.325) & 1.438  & 1.598 &-1.633 & -1.653  \\
			\hline
			  & $M_{Total},\mu_{B}$ &  & 2.736 & 0 & 0 & 0  \\ 
			\hline			  
			  & $E_{Total},Ry$ &  & -34847.62952446 & -34847.63240435 & -34847.63326628 & -34847.62949552 \\ 			  
			\hline			  
			  & $E-E_{FIM},mRy$ &  & 0 & -2.8799 & -3.7418 & 0.0289 \\ 			  			  
		\end{tabular}

\end{table}

\begin{table}[ht!]
	\begin{center}
		\caption{Dependence of effective interatomic exchange interactions(in meV) from magnetic structure.}\label{tab:table3}		
		\begin{tabular}{c|c|c|c|c|c|c|c|c|c|c|c|c} %
			$J_{ij}$ & FIM & AF1 & AF3 & AF4 \\
			\hline
			$J_{1}$     &-3.52  & 0    & 5.53 & 5.71 \\
			$J_{2}^{X}$ & 1.81  & 0    & 3.79 & 2.44 \\
			$J_{2}^{Y}$ & 1.81  & 0    & 2.98 & 2.37 \\
			$J_{3}^{X}$ & 4.22  & 0    & 4.77 & 5.36 \\
			$J_{3}^{Y}$ & 4.22  & 0    &-6.36 & 5.36 \\
			$J_{4}^{X}$ & 8.61  & 8.47 &10.13 &12.26 \\
			$J_{4}^{Y}$ & 8.61  & 8.47 & 8.66 & 9.26 \\
			$J_{5}$     & 0.66  & 2.21 & 1.84 &-1.84 \\
			$J_{6}$     &-0.32  & 5.67 & 5.18 & 4.20 \\
		\end{tabular}
	\end{center}
\end{table}

\newpage
\section{References}

\bibliography{biblio}
\bibliographystyle{ieeetr}

\end{document}